# Mapping ceramics research and its evolution

*Sylvain Deville[a] and Adam J. Stevenson*

*Laboratoire de Synthèse et Fonctionnalisation des Céramiques, UMR3080 CNRS/Saint-Gobain, Cavaillon F-84306, France,*

**Abstract**
We show here how a simple data mining of bibliographic records can be used to follow and help understand the evolution of a research domain, at a level that cannot be captured by reading individual papers in a field of this size. We illustrate the approach by investigating 43 years of research on ceramic materials, covered by 253k bibliographic records. The patterns of keywords used reveal the trends and the evolution of research ideas and priorities within the field. Simple, interactive tools based on co-word network analysis help us better appreciate the organization and relationships of ideas or individuals, and hopefully allow identification of unexplored concepts, connections, or approaches on a given topic.

**1. Introduction**
We have many tools that can be used to try to understand and follow both the history of and developments in our field. To date, scientists have relied primarily on the advice and guidance of mentors, textbooks, targeted searches of bibliographic databases and literature, attending conferences and workshops, and following a few targeted periodicals. However, with the ever-increasing array of sub-fields, conferences, published articles[1], and journals, it is increasingly difficult to keep tabs on even a small area of current research[2], let alone the historical developments that shaped the current environment. To fill this gap, a variety of new technologies, many social in nature, have been developed which merge the traditional discovery methods listed above with the social web: peer to peer interactions on the internet that aid discovery of new research and curation of information sources[3].

Along with the new social web technologies, other tools have emerged for obtaining, parsing, and understanding large text based data sets[4], and these techniques have been recently applied to understanding - on a coarse grained level - the development of scientific literature within various fields[5–9].

Here we use simple data mining to analyze the field of ceramics research, in a way that cannot be captured by reading individual papers, considering the large size of the field. We analyzed 253k bibliometric records spanning 43 years of ceramic research. This exploration allows us to identify the publishing habits within the field of ceramic research, the rise and fall of specific ceramic materials, topics, and techniques, and explore the relationships between academic research and technological changes. At a lower level, we illustrate how co-word and co-author network analysis can be a powerful tool for individual researchers to understand trends within their fields, identify potential collaborators, and help target their research.

---

[a] Corresponding author - Sylvain.Deville@saint-gobain.com (email), @DevilleSy (Twitter)

## 2. Rewinding the history of ceramic science

With bibliometric analysis, we can track changes to the entire field of ceramic research[b]. Here we illustrate several aspects of research: snapshots that capture the current research interest within a domain, and the evolution of specific ceramic materials, properties, characterization techniques, and applications revealed through the patterns of keywords usage.

### 2.1 Snapshots of current interests

A snapshot of the current interests of an entire field of research can be captured, for instance, using word clouds of keywords. In this representation, the size of each word is related to its statistical occurrence (in titles and abstracts) during the considered period. Although such analysis is very coarse, it can be used to reveal the dominant ideas, concepts, methods, or materials of interest at a given moment. In figure 1, examples are given for three different years: 1989 (6.1k records), 2000 (7k records), and 2011 (13.7k records).

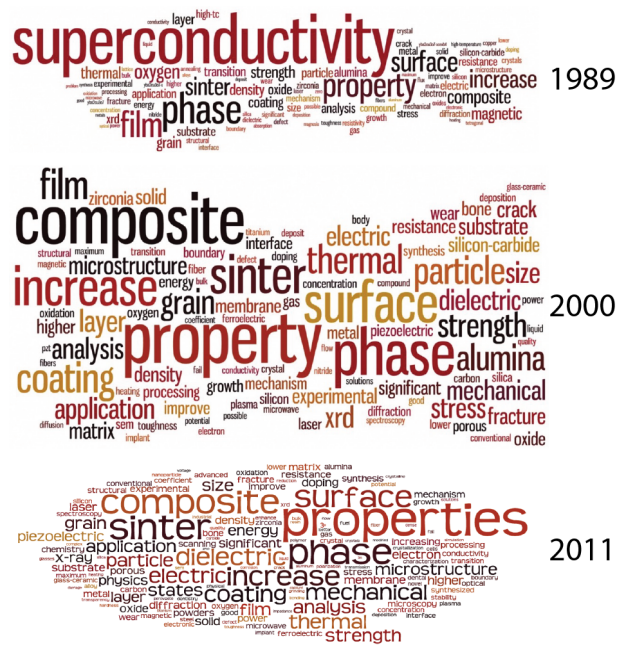

*Figure 1: Word cloud of most commonly used keywords in 1989, 2000, and 2011. The size of each word is related to its statistical occurrence in titles and abstracts during the given period.*

In 1989, the world of ceramic research was focused on exploring the superconductivity of ceramic materials. Three years before, Bednorz and Mueller revealed[10] that certain $ABO_3$ ceramic perovskite exhibit an unusually high critical temperature, and were later awarded

---

[b] We must be careful as the meaning of words may change and be plural. Words can fall in three categories: theoretical, methodological, and observational (Leydesdorff1997). This is potentially problematic if we are to look at co-words analysis (i.e., which words are commonly used together). Bearing in mind these limitations, we believe we can still use the bibliometric record to explore the evolution of the field and its characteristics. Here, we simply investigate the occurrences of words. In a technical field like ceramic research, this simpler analysis is less susceptible to words having several meanings.

with the Nobel prize for this discovery[c]. This discovery had potentially spectacular consequences, since having superconducting materials that operate at liquid nitrogen temperature is much more cost-effective than liquid helium temperature. The dominating keywords (*film*, *phase*, *surface*, *magnetic*, *high-tc*, *layer*, *oxygen*) are all related to the work on superconductivity. The impact of the work on superconductivity over the entire field of ceramic research can be observed at additional levels. The attention shifted for a while from the usual interests (as described below), but several other indicators reveal a change of pattern: the length of papers decreased (figure S2), as most papers were short communications reflecting a highly competitive and fast moving topic (the average length of superconductivity papers in 1988 is 3.5 pages, versus 5 pages for the average length of all papers), the average number of authors on a paper increased (figure S3) (as ceramists had to team up with physicists), and a spike in the number of papers (figure S1) and journals (figure S4) in which papers were published is observed. New journals were created to report specifically on this topic, such as *Superconductor Science and Technology* (est. 1988) or the *Journal of Superconductivity and Magnetism* (est. 1988)

In 2000, the situation was more balanced, but ceramic composites attract most of the attention. Finally, the 2011 word cloud is fairly homogeneous, reflecting the variety and richness of interests in ceramic research. Some concepts associated with ceramic materials still dominate (*sinter*, *phase*, *properties*, *microstructure*, *composite*), but a wider variety of keywords is observed: properties (*dielectric*, *electric*, *mechanical*), materials (*oxide*, *alumina*, *zirconia*, *carbon*), techniques (*analysis*, *x-ray*, *thermal*, *diffraction*), applications (*bone*, *coating*, *implant*), etc. This increase of diversity can also be seen as a consequence of the ever-increasing number of records every year (see fig S1) and the increase of multidisciplinary studies.

**2.2 Specific Interest for ceramic materials**
The corpus of records was obtained by searching for records with the keyword "*ceramic*". The field of ceramic research is, at least for the past 40 years, dominated by work on technical (as opposed to traditional) ceramics: the process, structure, properties, and applications of such materials. A statistical analysis (figure 2) of the patterns of keywords usage reveals the specific interest for the various ceramic materials, and captures their rise and fall.

Plots shown in figures 2 to 6 comprise two parts. The upper part shows the evolution of the relative occurrences of a few selected keywords. The occurrences have been normalized to account for the increasing number of records every year (see Methods section). The lower part reveals the relative evolution of occurrences for an individual keyword, normalized by the maximum number of occurrences for this specific keyword.

---

[c] The paper has now almost 13 000 citations as of today, and is probably one of the most cited papers in ceramic science.

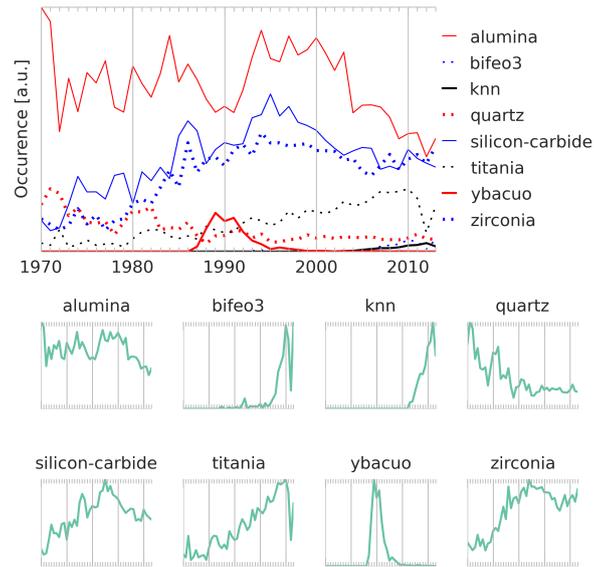

*Figure 2: Evolution the relative interest for various ceramic materials. Different fates can be observed: a gradual decrease of interest (quartz), a constant interest (alumina), a sudden burst (YBaCuO) which quickly disappears, a rising interest (titania), or the apparition of a new material (KNN).*

Ceramists will not be surprised that alumina is the ceramic material attracting most of the attention as it is both an industrially important structural material and a model material for fundamental studies. We can see how the past forty years have seen the attention shifting from structural and traditional to technical ceramics. Titania, for instance, which was investigated 40 years ago, has slowly and steadily become one of the most investigated ceramic materials. Quartz, on the other hand, was the second most investigated ceramic material in 1970 and is now the object of much less attention.

The appearance of novel materials can also be tracked, such as $Y_2Ba_3CuO_{7-d}$, KNN, or $BiFeO_3$, each of them appearing for different reasons, and having different fates. $Y_2Ba_3CuO_{7-d}$ was suddenly of interest when its superconductivity properties at liquid nitrogen temperatures were discovered[11]. The promises offered by high temperature superconducting ceramics, considered as a disruptive technology for electricity transmission, triggered a frenzy that lasted a few years, but faded quite rapidly when other physical limitations (critical current densities, internal heating, etc.) were shown to be as important to application as the critical temperature itself. The rise of KNN, a lead-free ceramic material, can be related to the international regulations on lead-free materials (in particular in Europe), for health and environmental reasons, which triggered work to find alternatives to lead[12], in particular in piezoelectric materials.

## 2.3 Properties

The evolution of specific properties can also be tracked (figure 3). Some historical properties of ceramics, such as mechanical strength, transparency, or piezoelectric properties have attracted a constant interest during the investigated period. In some cases, novel materials or concepts renewed previous interest (transparency). The properties related to nuclear applications have seen a spike of interest in the 80s (possibly as a consequence of the Three Mile Island nuclear accident in 1979) before falling to their lowest level today. The rising interest for dielectric applications can be linked to the rise of computers and more generally electronics and consumer electronics, where ceramic materials play a critical role. Superconductivity, as discussed previously, shows a unique pattern: an extremely rapid rise, until it became, by a large margin, the most investigated property for a few years, followed by a very rapid fall. Its influence is reflected in the sudden decrease observed, for instance, in mechanical and piezoelectric papers.

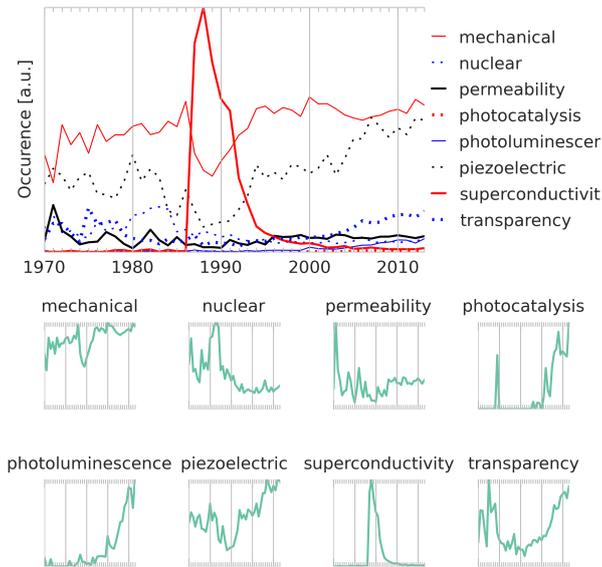

*Figure 3: Evolution of the relative interest for ceramic materials properties: mechanical, nuclear, permeability, photocatalysis, photoluminescence, piezoelectric, superconductivity, and transparency.*

## 2.4 The evolution of techniques and tools

Characterization techniques have always been and still are a key component of materials science and of ceramic science in particular. The statistical analysis of the corresponding keywords reveals the predominance of imaging techniques, such as SEM, TEM, or microscopy (figure 4). Since microstructural characterization is crucial to understanding both processing and properties, it is no surprise to see the predominance of these techniques. Their evolution also indicates different fates, depending on whether a useful technique becomes affordable, widely applicable, and easy enough for non-specialists to perform and analyze. Scanning electron microscopy (SEM) for instance is now routinely used in every materials science lab, and the price of a basic SEM is constantly decreasing. The use of SEM is therefore steadily increasing. Transmission electron microscopy is much less accessible, both from a technical (sample preparation, operation) and financial point of view. After its initial development where its use rapidly increased, its use has been fairly constant over the past 20 years. Alternatively, techniques that were previously standard are slowly becoming less and less relevant, such as optical microscopy, as other techniques are developed and become more accessible. Finally, the birth of techniques can also be tracked, such as the introduction of atomic force microscopy, focused ion beam microscopy, or confocal microscopy, although such a bibliometric analysis is not required to do it.

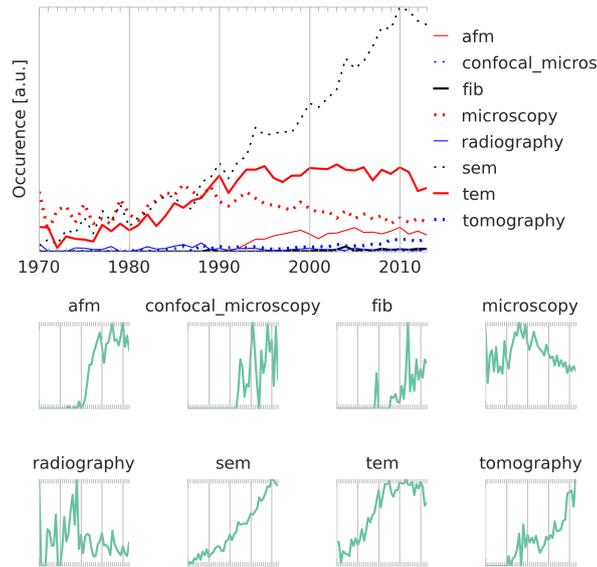

*Figure 4: Evolution the relative interest for characterization techniques*

**2.5 Industry and applications: driving forces for research?**
Application-related keywords (figure 5), can also reveal the driving forces for the work on ceramic materials. Historical applications of ceramic materials (*bearing*, *resistor*, *turbine*) have undergone a steady decline while other applications such as lasers, membranes, or solid oxide fuel cells increase in importance. Interestingly, ceramic materials were investigated for batteries applications before the interest faded. The last 15 years or so have seen a new rise of interest for this topic, due to both progress in ceramic materials and increasing energy considerations.

Industrialization may also have an impact on the number of publications in a specific area. For example, the number of records concerning turbine blades has decreased sharply since 1980, corresponding to the development (prior to 1980) and industrialization (post-1980) of thermal barrier coatings for turbine blades. This could be attributed to companies internalizing research in this area as it is industrialized, and these companies are less likely to publish their data than academic researchers that contributed heavily to the early development. Interestingly, we see some signs of a similar, but more recent, trend in solid oxide fuel cells. Beginning in 2000, there is a sharp increase in activity which stabilizes as the technology begins to mature from 2005-2010. In 2010, the publication activity begins to decrease. A possible explanation is that the technology has begun to be industrialized and is rapidly becoming commercially available. Thus, much of the research and development activity is moving inside of companies that are less likely to publish their results, following the pattern displayed by turbine blades. Observing the trend for superconductivity in Figure 3, we also see a sharp rise and subsequent fall in activity, but this trend cannot be related to commercial success. Instead, the sharp decline in activity surrounding high temperature superconductivity is most likely related to the fact that researchers quickly discovered several major and insurmountable challenges that prevented commercial adoption of the technology.

These three examples: turbine blades, solid oxide fuel cells (SOFC), and high temperature superconductivity are important for understanding how to use this historical literature data to target our research. By placing the research trends in context, we can better understand how or whether to enter certain fields. For example, to begin work with turbine blades, a field that has seen a substantial decrease in activity but has yielded many commercial products, a researcher should target industrial partners that have internalized much of the research and progress in this area over the last 30 years. For superconductivity, which follows a similar pattern but without commercial success, any new entrants into this field must clearly identify and have solutions to the fundamental problems that halted research in this area in the early 90s. With these two examples in mind, a researcher looking at these trends and with a desire to work with SOFCs must carefully consider whether the recent decline in SOFC publishing activity is a statistical oddity, a result of increasing commercialization that will necessitate industrial partnerships to develop research programs in this area, or whether it is a result of fundamental physical limits that must be solved for the technology to be developed further.

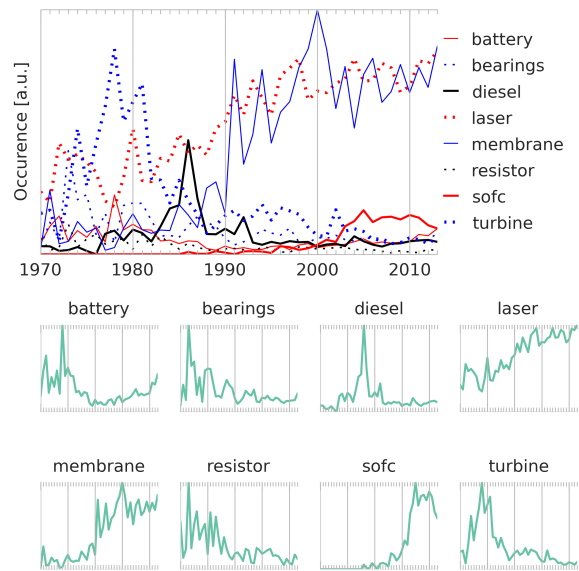

*Figure 5: Evolution the relative interest for application-related keywords*

## 2.6 Writing style and the positive bias

It is an unfortunate habit to report only positive results[d], which are supposedly more interesting than failures. Scientific writing is also supposed to be objective and unbiased. This intuitively strong bias towards positive results[13] is revealed here by the statistical analysis of the words used in the titles and abstracts of the papers (figure 6). Keywords with a positive connotation, such as *enhance*, *improve*, *innovative*, *remarkable*, *significant*, *successful*, or *promising* have been increasingly used over the past forty years. Trying to attract attention, funding, and fame, the use of hype keywords is apparently ever increasing (although most journal editors do not encourage it), with no sign of stopping or slowing down anytime soon, although, in a interesting twist, *new* (new results, new materials, new mechanisms, new processes) show a relatively steady decrease.

---

[d] with few exceptions such as the journal of negative results http://www.jnr-eeg.org , in ecology and evolutionary biology.

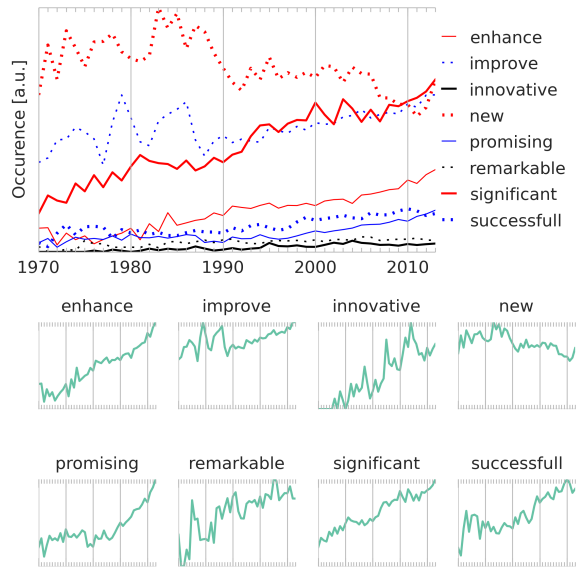

*Figure 6: Evolution of the use of keywords with a positive bias.*

**3. Networks: linking concepts, understanding the organization of research**

Another interesting use of data mining bibliographic records is the co-word network analysis. Two examples are described here: co-word (keywords) and co-authors (individuals) analysis, to illustrate how we can take advantage of such a tool.

The individuals' network was built from the author list on each record. Relationships (*links*) are established between co-authors of the same paper. No preferences were given to the first and to the corresponding author in this analysis, so that each author has the same weight on the network. The network obtained (figure 7) shows clusters of individuals. The size of each bubble is given by its statistical importance: individuals appearing more often in records appear as a bubble of a greater size. Bubbles of the same color belong to the same network. Such a network is a very convenient way to identify the key players in a specific domain, and reveal different organizational patterns depending on the domain. A corpus of records was obtained by running a search with two different keywords: transparent ceramic (4.4k records), and solar cells (98k records).

Several observations can be made on the characteristics and topology of the network. First, individual groups can easily be observed on the network, with two types of organizations: collaborative groups, where all the bubbles of a given group have the same size, and groups with a clear leader, having a much larger size than its co-authors. Although not shown on the figure, the name of each person is attached to each bubble, so that the most active individuals in a given domain can be rapidly identified.

These networks also reveal different habits depending on the domain of interest. For transparent ceramics, very few links exist between the different groups. Work within this field can be divided up into materials fabrication, optical properties measurements, and other (mechanical, etc.) properties measurements. Since many research organizations have groups that can fulfill these tasks, many small collaborative groups appear. Once a functional group is established, this group can work on many different materials and applications with little need to move outside the network. The network on solar cells shows a very different topology, with a much greater number of links between the various groups. This is possibly due to the complexity required to process, characterize, and analyze the performances of solar cells on multiple different levels from materials, to cells, to modules. A single group or even a single research center may be less likely to have all the required expertise. In particular, standard efficiency testing for solar cells can be performed in relatively few locations, and these sites act as links between various groups doing materials, cell, and module development.

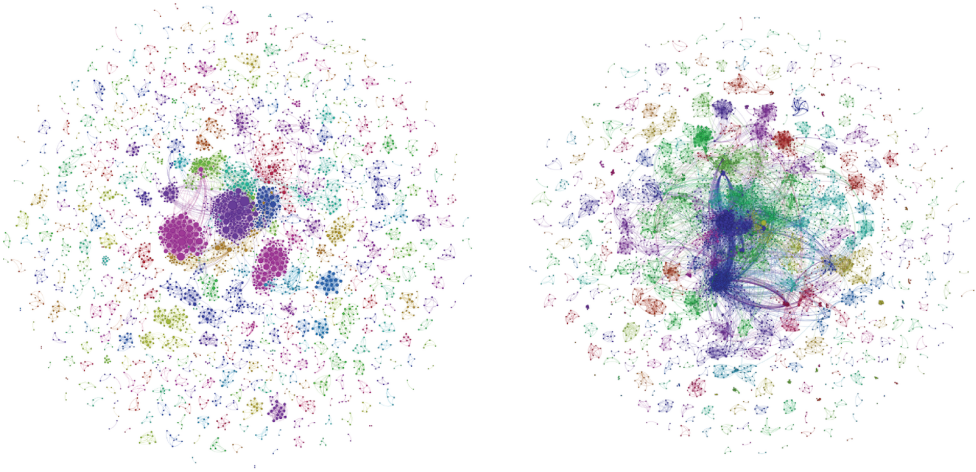

*Figure 7: Network of individuals with records on transparent ceramics (left) and solar cells (right).*

Such analysis is limited by the difficulty to address author's ambiguity and therefore giving correct credits to individuals. This problem is particularly prominent in countries such as China or Korea, where the same family name is shared by a very large number of individuals. Initiatives like ORCID should hopefully provide a solution for this, once (if?) they become widely adopted.

A similar network can be built if we analyze the keywords attached to the records. Links are established between keywords listed in the same paper. We then performed a class analysis: keywords sharing the same relationships belong to the same class, and are represented with the same color. This clustering allows identifying the organization of concepts and ideas within a given field. Similar to the individuals' network discussed previously, the bubble size of each keyword is related to its statistical occurrence. For the sake of clarity, only the keywords with the highest occurrence are shown (*figure 8*).

Although we can obtain the keyword network for the entire corpus of records investigated here, a printed figure is not the ideal representation for such a complex and detailed network. To illustrate the interest of such a tool, we will discuss the keyword network obtained for records related to transparent ceramics.

The main groups of concepts can be identified by their color: properties (light blue in the lower part: *optical*, *electric*, *dielectric*), materials (green on the right hand side: *compound*, *carbide*, *sodium*, *aluminum*, *yttrium*), processing methods (light green in upper part: *method*, *chemical*, *deposition*, *sol*, *gel*), densification methods (brown in upper part: *sinter*, *plasma*, *hot pressing*), microstructure (violet in lower part: *structure*, *particle*, *growth*, *grain*, *pore*, *crack*), etc. Such representation can be particularly useful, for instance, to identify the main materials or techniques investigated in a given domain. They may also be used to identify missing links and make new connections.

## 4. Conclusions

We illustrated here how a simple data mining can be a powerful tool to map a field of research and explore the relationships between academic research and technological changes. Such analysis can be done in real time, and eventually at a finer scale that shown here. Networks revealing the relationships between concepts can be built to understand the organization of research and possibly identify new, unexplored connections.

We exposed here a simple set of tools: bibliographic records, which are now widely available in machine-readable format, a script, and open-source software to build, represent, and interact with the network. Extensive computing resources are not required: the analysis was ran on a standard laptop. We can make the analysis a lot more complex, in particular if data mining of the full-text of records is available. We believe that such approach is complementary to the usual toolbox used in research. By offering a different look at the records, it may help discovering hidden connections and unexplored approaches.

*Figure 8: Network of keywords (co-word analysis) related to transparent ceramics records. A wider link indicates a stronger occurrence of the two words in the same record (e.g. grain/size, thin/film, solid/state, rare/earth).*

**Acknowledgements**
We acknowledge Stuart Cantrill (@stuartcantrill), Alexis Verger (@Alexis_Verger), and Benjamin Abecassis (@b_abk6) for the pre-publication peer-review of the paper through twitter.

**Methods**

The dataset was constructed from the Scopus database, using the simple search query "*ceramic*" in title, abstract, or source title. After duplicates were removed, we were left with 253k unique records over 43 years. Records include articles (the vast majority), conference papers, review, book, and letters. Parsing and analysis of the records were made with a python script. We used a stop list to remove common words (*the*, *and*, *with*) and numbers, pronouns, conjugations of common words (*have*, *will*, *can*). Similar words have been merged (singular/plural, comparative and superlative, conjugations of verbs). Specialized nouns and adjectives (e.g. *pore*, *porous* and *porosity*) were equated. We then performed a co-word (words that frequently appear together) and co-authors (author that frequently appear together in a record) analysis.

With 619 records in 1970 and 13.5k records in 2013, there has been a constant increase in the number of bibliographic records over the past 40 years. At the same time, the average length of the records title also increased (figure S2 in SOM). Normalization was therefore performed by the relative volume of data (raw length of records list for the year). A second normalization was performed for individual keyword (subpanels in figures 2 to 6) by the maximum count of occurrences of the individual keyword, to explore their relative evolution and identify rapidly emerging or declining concepts.

The word clouds were obtained using a freely available online tool (http://wordle.net). The networks were built with Gephi (https://gephi.org/), an open source software.

**Supplementary Materials**

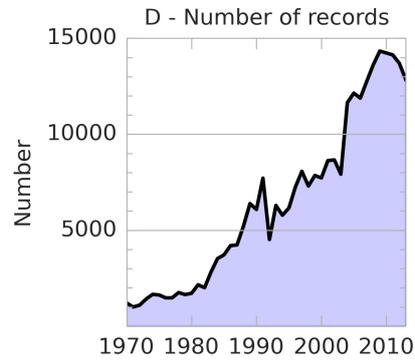

Figure S1: Number of records per year, since 1970.

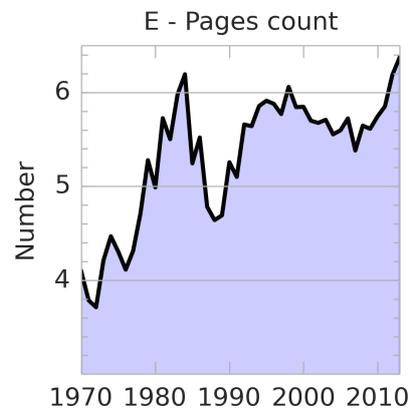

Figure S2: Average page count per record, since 1970.

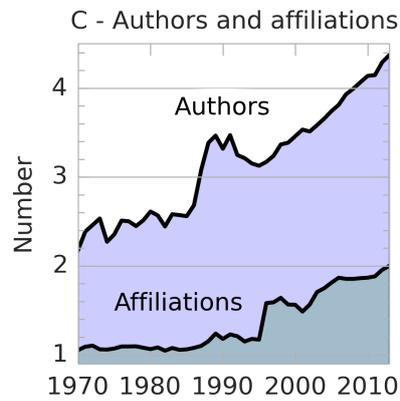

Figure S3: Average number of authors and affiliations per paper, since 1970.

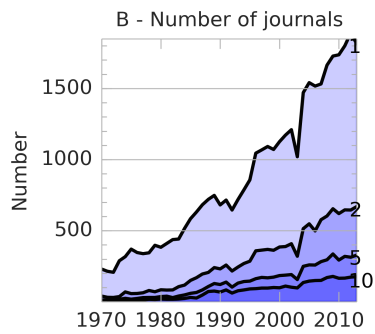

Figure S4: Average number of journals in which records were found every year, since 1970. The 1, 2, 5 and 10 figures correspond to the number of journals which published at least 1, 2, 5 and 10 papers per year on the topic of ceramic research.